\documentclass[journal,comsoc]{IEEEtran}
\IEEEoverridecommandlockouts
\usepackage{cite}
\usepackage{amsmath,amssymb,amsfonts}
\usepackage{algorithmic}
\usepackage{graphicx}
\usepackage{subcaption}
\usepackage{textcomp}
\usepackage{url}
\usepackage{tabularx}
\usepackage{array}

\newcolumntype{Y}{>{\hsize=1.3\hsize}X}
\newcolumntype{Z}{>{\hsize=0.7\hsize}X}

\usepackage{placeins} 
\usepackage{afterpage} 

\def\BibTeX{{\rm B\kern-.05em{\sc i\kern-.025em b}\kern-.08em
    T\kern-.1667em\lower.7ex\hbox{E}\kern-.125emX}}
\begin{document}

\title{Vehicle-to-Grid as a 5G Smart Grid Vertical: Non-Technical Barriers and Implications for Communication Networks\\}

\author{
    \IEEEauthorblockN{
        Shangqing Wang\IEEEauthorrefmark{1}, 
        Laura del Rio Carazo\IEEEauthorrefmark{3}, 
        and Frank H. P. Fitzek\IEEEauthorrefmark{1}\IEEEauthorrefmark{2}
    }

    \IEEEauthorblockA{
        \IEEEauthorrefmark{1} Deutsche Telekom Chair of Communication Networks, Technische Universitaet, Germany
    }
    
    \IEEEauthorblockA{
        \IEEEauthorrefmark{2} Centre for Tactile Internet with Human-in-the-Loop (CeTI), Technische Universitaet, Germany
    }

    \IEEEauthorblockA{
        \IEEEauthorrefmark{3} Escuela Técnica Superior de Ingenieros de Telecomunicación,\\
        Universidad Politécnica de Madrid (UPM), Spain
    }

    \IEEEauthorblockA{
        E-mails: shangqing.wang@tu-dresden.de, laura.delrio@upm.es, frank.fitzek@tu-dresden.de
    }
}

\maketitle

\bstctlcite{bibliography:BSTcontrol} 

\begin{abstract}
Vehicle-to-Grid (V2G) and broader Vehicle-to-Everything (V2X) technologies are technically mature and widely demonstrated, yet large-scale deployment is constrained mainly by non-technical rather than communication or power-electronics limits. This paper targets the wireless communications community and frames V2G as a 5G-enabled smart grid vertical, linking business, governance, social, and infrastructure barriers to concrete communication-system requirements. Building on a PRISMA-guided systematic review of 974 V2G/V2X publications (2009--2025), and 162 implementation-critical studies, we adopt a four-domain framework of non-technical barriers: Business/Economic, Governance/Policy, Social, and Infrastructure/Ecosystem. Temporal and regional analyses show a shift from technical dominance to multidisciplinary integration after 2021. We translate these domains into communication requirements for V2G verticals, including fine-grained metering and settlement, protocol interoperability (e.g., ISO~15118, OCPP), privacy-by-design data governance, latency- and reliability-differentiated services, and edge--cloud partitioning for flexibility control. The results demonstrate that 5G design for V2G cannot be a purely technical optimization task and must integrate socio-technical constraints from the outset, suggesting research directions for sustainable, data-driven V2G communication architectures.
\end{abstract}


\section{Introduction}
Vehicle-to-Grid (V2G) and broader Vehicle-to-Everything (V2X) technologies have emerged as key enablers of low-carbon power systems and smart cities, allowing electric vehicles not only to decarbonize transport but also to provide flexibility and storage to electricity networks facing high renewable penetration and evolving resiliency demands~\cite{kumar2024comprehensive,mobilityhouse2024v2g}. Although considerable technical progress has been achieved in device interoperability, communication protocols, bidirectional charging, and battery management, large-scale deployment remains constrained less by engineering limits than by non-technical challenges such as business feasibility, regulatory readiness, user acceptance, and ecosystem coordination~\cite{ffe2025unlockingV2G,rao2025vehicle,wang2024bidirectional}.

\textbf{Why now?} With 974 V2G/V2X studies published between 2009 and 2025, technical solutions abound, yet deployment trajectories are fragmented and often stall after pilots, indicating that non-technical barriers now dominate the bottleneck. Analyzing these factors is timely as Europe commercializes V2G mainly through pilots, several Asian countries focus on infrastructure scaling, and global standards mature, revealing implementation gaps that this line of work addresses~\cite{kumar2024comprehensive,ffe2025unlockingV2G,rao2025vehicle}. In parallel, V2G is increasingly positioned as a 5G-enabled smart grid vertical within IoT-based energy and mobility systems, making it essential to understand how socio-technical barriers translate into communication-system requirements.

This article builds on a companion systematic review published in Energies~\cite{wang2026nontechnical}, which develops a four-domain socio-technical framework, transition levers, and non-technical key performance indicators (KPIs) for V2G deployment based on a PRISMA-guided analysis of 974 studies, 162 implementation-critical articles, and a 95-paper deep-dive knowledge base. In that work, the focus lies on business–economic, governance–policy, social, and infrastructure and ecosystem dimensions of V2G transitions; in this article, the same corpus and framework are reused and extended to derive communication-system requirements and 5G design implications for V2G as a smart grid vertical.

The article targets the communications and networking community and examines how non-technical constraints identified in the socio-technical review shape concrete design choices for V2G communication architectures. Building on the full 974-article corpus and the classified subset of implementation-critical studies, the analysis focuses on how business models, regulatory structures, user practices, and infrastructure readiness map to requirements on metering and settlement, protocol interoperability, privacy-by-design data governance, latency and reliability classes, and edge–cloud partitioning for flexibility control~\cite{wang2024bidirectional,wang2025realworld}. Recent work on bidirectional charging architectures and communication interfaces between electric vehicles, charging infrastructure, and optimization backends further illustrates how real-world V2G deployments depend on robust, low-latency, and interoperable communication systems in concrete urban districts and 5G-enabled smart city pilots~\cite{wang2025realworld,wang2025simulationostra,wang2024mobilities5g,wang2024bidirectionalusecases}.

\textbf{The main contributions of this article are as follows:}
\begin{enumerate}
  \item \textbf{A communication-focused synthesis of a PRISMA-screened V2G/V2X corpus.} Building on the Energies companion review, this article uses the same 974-study dataset and classified subset of 162 implementation-critical articles as a basis for deriving communication-layer implications for V2G verticals, rather than re-presenting the full socio-technical analysis.
  \item \textbf{A mapping from four non-technical domains to V2G-specific communication requirements.} The article systematically links Business/Economic, Governance/Policy, Social, and Infrastructure \& Ecosystem domains to practical requirements on metering and settlement support, protocol interoperability (e.g., ISO~15118 and OCPP), privacy and user-control mechanisms, and differentiated latency and reliability classes for V2G services.
  \item \textbf{Semi-quantitative 5G requirements and slicing guidelines for V2G.} Based on service characteristics and non-technical constraints, the paper proposes indicative latency, reliability, and availability targets for V2G sub-services (such as fast ancillary services versus slower demand response), and discusses how these differ from generic V2X requirements in terms of network slicing and service orchestration.
  \item \textbf{Identification of research directions and standards gaps in V2G communication architectures.} The article highlights open issues in aligning existing 5G standards and implementations with V2G-derived requirements, including Quality of Service (QoS) exposure for energy services, integration of V2G-relevant application protocols into network policy control, and opportunities for digital-twin support, advanced optimization, and secure data governance in future 5G-based V2G deployments.
\end{enumerate}

\section{Methodology}
\label{sec:methodology}
This article reuses the PRISMA-guided systematic review protocol developed in a companion Energies article on non-technical V2G barriers and transition pathways, which documents the full search strings, screening flow, coding rubric, and implementation-critical corpus in detail~\cite{wang2026nontechnical,page2021prisma}. Building on that foundation, the present work focuses on how the resulting socio-technical framework and classified studies inform communication-system requirements for V2G as a 5G smart grid vertical. For completeness, this section summarizes the key elements relevant to the communication-layer analysis: (i) research questions and search strategy; (ii) multi-stage screening and inclusion criteria; and (iii) coding of studies into technical, non-technical, and mixed categories across four non-technical domains.

\subsection{Research questions and search strategy}
The original review is guided by three research questions~\cite{wang2026nontechnical}: (RQ1) What are the barriers and enablers to V2G technology transfer across contexts? (RQ2) Does the field disproportionately emphasize technical challenges over governance, market, behavioural, or policy issues? (RQ3) Which non-technical obstacles—such as policy gaps, business model uncertainty, user resistance, regulation, and stakeholder misalignment—remain most overlooked?

The primary search was conducted in the Web of Science Core Collection for the period 2009--2025, capturing the evolution from early V2G concepts to recent deployment efforts~\cite{wang2026nontechnical}. Boolean keyword logic combined V2G/V2X terms with implementation- and barrier-related concepts, for example:
\emph{(``V2G'' OR ``bidirectional charging'' OR ``vehicle-to-grid'') AND (``technology transfer'' OR ``barrier'' OR ``policy'' OR ``market'' OR ``governance'' OR ``user acceptance'' OR ``stakeholders'')}.
The initial search yielded 974 records.

\subsection{Screening, inclusion, and coding}
Screening proceeded in two phases using a keyword-based protocol~\cite{wang2026nontechnical}. In Phase~1 (title–abstract filtering), titles and abstracts were scanned for implementation signals and barrier-related terms (e.g., ``pilot'', ``demo'', ``implementation'', ``barrier'', ``policy'', ``market'', ``adoption''), reducing the set to 788 records. In Phase~2 (prioritization and abstract scoring), each abstract was assigned a relevance score between~0 and~3 based on the density of the same keyword list; studies with scores $\ge 3$ advanced to full-text review, yielding 164 deployment-relevant papers. After full-text assessment, 162 articles were retained as implementation-critical V2G/V2X studies.

Inclusion criteria required peer-reviewed journal articles and reviews that (i) address V2G or broader V2X (including V2H, V2B, V2L) in a practical or implementation-relevant setting (pilots, field trials, empirical case studies, or realistic market/system modelling), and (ii) discuss at least one dimension of technology transfer (technical or non-technical), such as deployment barriers, business models, regulatory issues, user behaviour, or infrastructure readiness~\cite{wang2026nontechnical}. Exclusion criteria removed non-peer-reviewed items, generic EV charging or storage studies without explicit V2G/V2X focus, and purely theoretical or simulation-only works without links to deployment contexts.

Full-text coding assigned each of the 162 implementation-critical articles a principal tag—technical, non-technical, or mixed—based on standardized definitions shared with the Energies review~\cite{wang2026nontechnical}. Technical studies focused on engineering components (protocols, grid stability, algorithmic control, power electronics, communication architectures); non-technical studies addressed policy, markets, stakeholder engagement, incentives, finance, and social acceptance; mixed studies treated both dimensions substantively. Non-technical and mixed studies were further mapped into one or more of four recurrent non-technical domains (Business/Economic, Governance/Policy, Social, Infrastructure \& Ecosystem), forming the socio-technical framework used in this article to derive communication requirements.

Table~\ref{tab:summary} summarizes the final classification, which is identical to that reported in~\cite{wang2026nontechnical}. In subsequent sections, the communication-system analysis focuses particularly on the 95 studies (all 44 non-technical and 51 mixed) in which non-technical aspects dominate, as these constitute the primary technology-transfer knowledge base for V2G.

\begin{table}[t]
\centering
\caption{Classification of 162 implementation-critical V2G/V2X articles.}
\begin{tabular}{l c}
\hline
\textbf{Category} & \textbf{Article count} \\
\hline
Technical     & 67 \\
Non-Technical & 44 \\
Mixed         & 51 \\
\hline
\end{tabular}
\label{tab:summary}
\end{table}

\section{Overview of V2G research trends}
\label{sec:results}

\subsection{Scope, regional distribution, and key barriers}
This systematic review covers 974 V2G/V2X articles published between 2009 and 2025, demonstrating sustained global interest in implementation across diverse regions. The corpus spans Europe, Asia/Middle East, the Americas, Australia, Africa, and global reviews, with clear regional emphases.\cite{weiller2014using,mwasilu2014electric,hosseini2014survey,sovacool2009beyond,song2010integration} Europe focuses on business models, taxation, stakeholder engagement, and smart charging tariffs; Asia and the Middle East emphasize megacity infrastructure, business ecosystems, trading mechanisms, and CHAdeMO-based PV–V2G decarbonization; the Americas highlight frequency regulation, resilience, and fleet electrification; Australia analyses consumer barriers and long-distance travel patterns; Africa concentrates on load-shedding mitigation and minibus fleet electrification; and global reviews benchmark feasibility, economic viability, and policy barriers~\cite{de2024economic,sabadini2025does,zhang2024shanghai,garofalaki2022electric,philip2023adoption,abdelfattah2024artificial,mojumder2022electric}. 

Table~\ref{tab:regional_compact} summarizes these patterns and shows how regulatory maturity, market design, user behaviour, and infrastructure readiness jointly shape V2G trajectories, which in turn impose different requirements on communication networks and service architectures.

\begin{table*}[t]
\centering
\caption{Regional V2G research foci and typical non-technical barriers (2009--2025).}
\label{tab:regional_compact}
\begin{tabular}{p{1.8cm} p{4.0cm} p{7.0cm}}
\hline
\textbf{Region} & \textbf{Main focus} & \textbf{Typical non-technical barriers} \\
\hline
Europe & Business models, tariffs, pilots, planning with RES & Complex regulation and taxation, interoperability, pilot scalability, user trust, coordination among actors~\cite{de2024economic,sabadini2025does,weiller2014using} \\
Asia / Middle East & Megacity infrastructure, business ecosystems, trading, CHAdeMO & Fragmented standards, uneven grid modernization, uncertain revenues, policy delays, coordination across actors~\cite{zhang2024shanghai,hasan2023critical,karim2024impact} \\
Americas & Frequency regulation, resilience, fleets, techno-economics & Interconnection constraints, slow market reforms, limited bidirectional EV availability, inconsistent policies~\cite{shokrzadeh2017simplified,garofalaki2022electric,bitencourt2025comprehensive} \\
Australia & Adoption lag, participation, remote grids & Low EV penetration, long-distance travel, limited incentives, sparse infrastructure~\cite{philip2023adoption,lucas2024participation} \\
Africa & Load-shedding, minibus fleets, grid support & Affordability, limited charging, grid instability, weak market mechanisms~\cite{abdelfattah2024artificial,ahjum2023vehicle} \\
Global & Cross-regional reviews, policy and market comparisons & Lack of unified protocols, insufficient real-world pilots, unclear revenue guarantees~\cite{mojumder2022electric,rehman2023comprehensive} \\
\hline
\end{tabular}
\end{table*}

\subsection{Thematic evolution over time}
Thematic analysis of the 974 studies reveals a consistent evolution from conceptual discussions to deployment-oriented work. Early publications (2009--2011) established the feasibility of using EVs as flexible storage and highlighted socio-technical obstacles and grid-integration challenges.\cite{sovacool2009beyond,markel2009communication,song2010integration} Between 2012 and 2016, research moved towards detailed technical modelling of smart grid integration, ancillary services, and optimization of bidirectional converters and charging strategies, while beginning to acknowledge privacy, business, and investment barriers~\cite{mwasilu2014electric,hosseini2014survey,mohamed2015power,guo2016potential}.

From 2017 to 2021, attention expanded to economic valuation, policy mechanisms, and the application of AI/ML for V2G scheduling and system planning, alongside emerging work on regulatory frameworks and business models~\cite{wei2017influence,kester2018promoting,habib2018assessment,mao2019intelligent,wang2019distribution}. Post‑2021, the literature exhibits a ``hyper‑specialization'' phase, with rapidly growing numbers of studies on cybersecurity, blockchain-based coordination, user acceptance, digital twins, and large-scale planning for extreme fast charging and resilience~\cite{islam2022state,garofalaki2022electric,eltamaly2023smart,bakhuis2025exploring,saba2025digital}.

Overall, this trajectory points to four broad phases: (i) conceptual and feasibility-oriented work; (ii) technical modelling and V2X expansion; (iii) increased use of advanced optimization and machine-learning techniques in V2G scheduling and system planning; and (iv) commercial viability, governance, and user-centric concerns. These shifts motivate the need to move beyond purely technical assessments and to structure non-technical deployment barriers into a coherent framework, which we develop in the next section.

\section{Four-domain framework of non-technical barriers}
\label{sec:framework}

\begin{figure*}[t]
\centering
\includegraphics[width=1.0\textwidth]{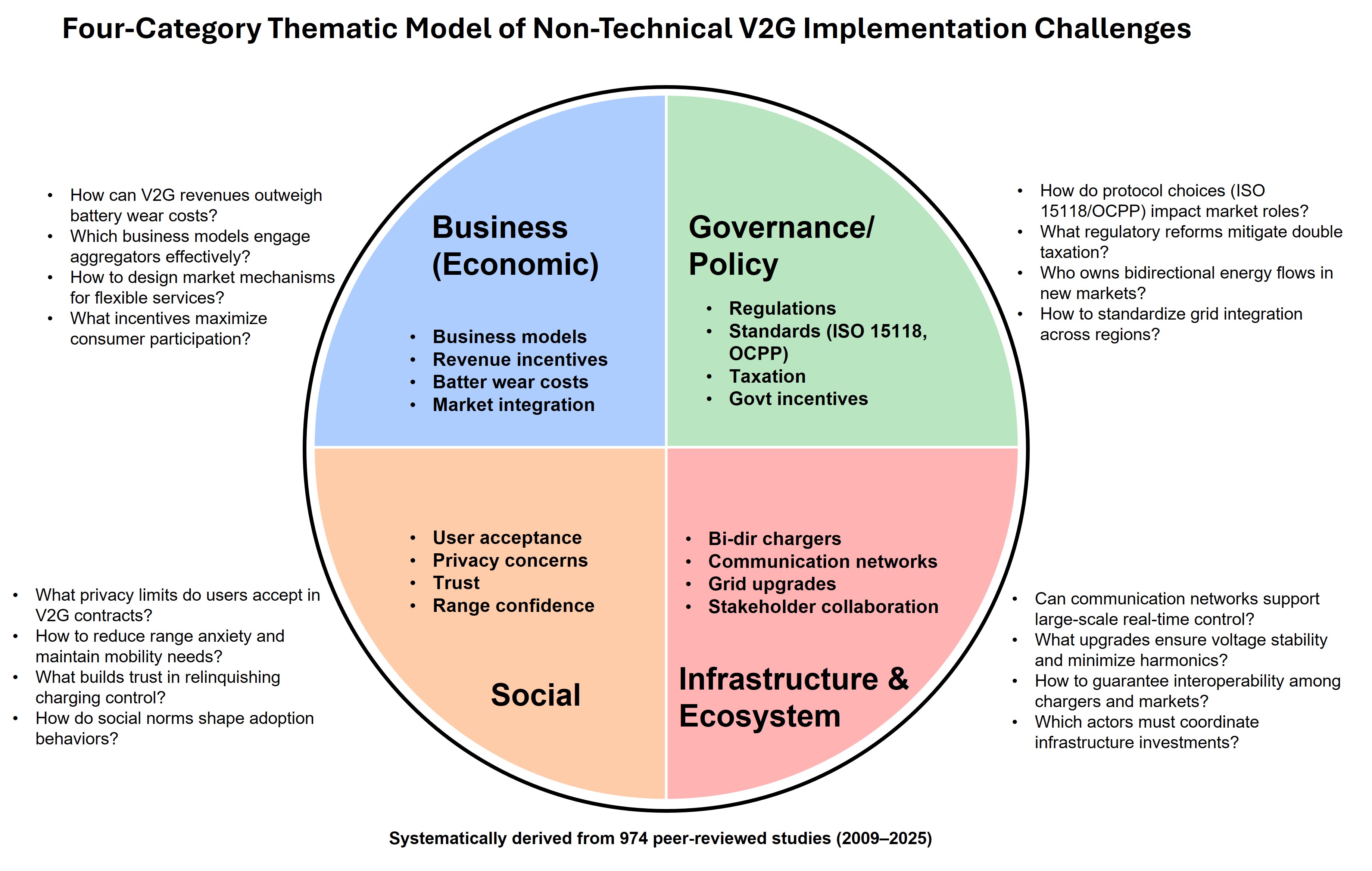}
\caption{Four-domain framework of non-technical factors shaping V2G implementation. Each quadrant shows one main category (inner layer), representative subcategories (middle), and strategic questions (outer), derived from 974 studies (2009--2025). The framework provides a structured basis for deriving 5G-enabled V2G communication and control requirements.}
\label{fig:framework_overview}
\end{figure*}

Figure~\ref{fig:framework_overview} summarizes four non-technical domains that emerged from coding the 974-article corpus: Business (Economic), Governance/Policy, Social, and Infrastructure \& Ecosystem~\cite{mojumder2022electric,weiller2014using,de2024economic,sabadini2025does}. Each domain encapsulates recurring implementation barriers and enablers and, taken together, provides a structured lens for deriving requirements for 5G-enabled V2G services. Representative questions include, under Governance/Policy, \emph{``How do ISO~15118/OCPP protocol choices impact emerging market roles?''} and, within Social, \emph{``Which data exposure levels are acceptable to EV users under different contracts?''}~\cite{lekidis2024mobility,ascher2024data,zhang2022toward}

\subsection{Business (Economic)}

V2G business models remain fragile, with studies stressing missing multi-stakeholder revenue models, weak incentives for EV users, and uncertain profitability once infrastructure and battery wear are accounted for~\cite{mojumder2022electric,weiller2014using,huang2021electric,sabadini2025does,qi2025optimizing}. Mechanisms for integrating V2G into existing markets (frequency regulation, demand response, arbitrage) are still immature~\cite{tan2016integration}.

\textbf{Implications for communication networks:} platforms must support fine-grained metering and settlement, secure aggregator–user transactions, and low-latency exchange of tariff and market signals so that V2G services can be priced and settled in near real time~\cite{wan2024feasibility,wang2024v2g}.

\subsection{Governance/Policy}

Governance and policy work highlights fragmented or unclear regulatory frameworks, the slow recognition of aggregators as market actors, and persistent uncertainty around responsibilities and taxation~\cite{tirunagari2022reaping,venegas2021active,sabadini2025does,kester2018promoting}. At the technical level, standardization of bidirectional charging and communication protocols (ISO~15118, OCPP) is a recurring prerequisite for scalable deployment~\cite{lekidis2024mobility,Rodriguez2021,tasnim2024critical,wellisch2015vehicle}.

\textbf{Implications for communication networks:} V2G service stacks must embed protocol compliance checks, role-based authentication, and standardized data schemas for regulatory reporting, allowing regulators and system operators to audit behaviour without compromising privacy~\cite{ascher2024data,shahed2024battery}.

\subsection{Social}

Social studies emphasize that user behaviour, perceptions, and trust strongly condition V2G uptake. Key concerns include battery degradation, range anxiety, data privacy, and loss of control over charging, alongside a lack of clear, accessible communication about benefits and risks~\cite{bakhuis2025exploring,neaimeh2025learning,lucas2024participation,lehtola2025vehicle,krueger2020integration,zhang2022toward,razzaque2025cybersecurity}.

\textbf{Implications for communication networks:} communication architectures should provide granular privacy and consent mechanisms, user‑configurable participation profiles, transparent explanations of control actions, and override options that still preserve grid service quality~\cite{shang2024information,wang2024charging}.

\subsection{Infrastructure \& Ecosystem}

Infrastructure and ecosystem work shows that V2G deployment depends on reliable bidirectional chargers, secure and low-latency communication networks for coordinating EV fleets, grid readiness for bidirectional flows, and coordinated investment across multiple stakeholders~\cite{lekidis2024mobility,tasnim2024critical,mlindelwa2024overview,jiang2016secure,al2016wave,cupan2024technical,ibrahim2024analysis,weiller2014using,wang2024bidirectional}.

\textbf{Implications for communication networks:} future 5G-based V2G systems must support scalable, interoperable protocols and resilient control architectures, with clear latency and reliability classes (e.g., ultra-fast services vs. slower demand response) and tight integration with grid monitoring and edge–cloud control~\cite{tao2017foud,meraj2025bidirectional,razzaque2025cybersecurity}.

\medskip
In aggregate, these four domains capture the main non-technical constraints that V2G deployments face worldwide and indicate where communication system design must align with business, regulatory, social, and infrastructure realities. The next section translates this framework into concrete implications for 5G communication architectures and services.

\section{Communication Requirements and 5G Architectures for V2G}
\label{sec:comm-req}

Building on the four-domain socio-technical framework and the implementation-critical corpus described in the companion Energies review~\cite{wang2026nontechnical}, this section derives indicative communication requirements for V2G services and outlines architectural patterns for 5G-enabled V2G verticals. The analysis is grounded both in the literature on V2G implementation barriers and in a private-5G bidirectional charging pilot in Dresden's Ostra district, where a layered architecture with an Integrated Mobility and Energy Platform (IMEP) and a Charging Optimization System (COS) has been deployed~\cite{Wang2026:Bridging}.

\subsection{Service-level communication requirements}
\label{subsec:service-qos}

The implementation-focused studies and the Dresden pilot indicate that V2G deployments typically bundle several sub-services with distinct temporal and reliability characteristics, such as fast grid support, congestion management, market-oriented scheduling, and user-facing applications~\cite{wang2026nontechnical,Wang2026:Bridging}. Table~\ref{tab:qos} summarizes indicative communication requirements for four representative sub-services, derived from the service characteristics reported in the corpus and from the IMEP/COS architecture.

\begin{table*}[t]
\centering
\caption{Indicative communication requirements for representative V2G sub-services. Values are derived from implementation-focused V2G studies and the Dresden Ostra pilot.}
\label{tab:qos}
\begin{tabular}{p{3.5cm} p{2.0cm} p{2.0cm} p{5.0cm}}
\hline
\textbf{Sub-service} & \textbf{Latency} & \textbf{Reliability} & \textbf{Typical slice profile} \\
\hline
Fast grid support (e.g., frequency response, fast active power adjustment) & $< 20$ ms & Very high & URLLC-like control slice for critical grid-support traffic \\
Distribution-level congestion management and voltage control & $0.5$--$5$ s & High & mMTC/eMBB mix for telemetry and control at feeder level \\
Scheduling and settlement (day-ahead / intraday participation, billing, auditing) & $10$--$60$ s & High & eMBB / best-effort slice with strong integrity and retention requirements \\
User-facing applications (notifications, consent updates, participation control) & Seconds & Medium--high & eMBB / best-effort slice with strong privacy and access control \\
\hline
\end{tabular}
\end{table*}
The fast grid-support row reflects ancillary service use cases in the Americas and Europe where V2G is integrated into frequency regulation and fast flexibility products; here, latency targets in the tens of milliseconds are needed for EV coordination to contribute meaningfully to system stability, aligning with the private-5G deployment in Dresden, which operates in the 3.9~GHz band with 100~MHz bandwidth and a target latency below 15~ms for critical control flows~\cite{bitencourt2025comprehensive,Wang2026:Bridging}. Distribution-level congestion management and voltage control require slower but still responsive coordination (sub-second to a few seconds), whereas market scheduling and settlement primarily demand high integrity, availability, and traceability rather than ultra-low latency~\cite{wang2026nontechnical}. User-facing applications can tolerate higher latency but must be supported by architectures that guarantee granular privacy and transparent control, as highlighted by social-domain studies~\cite{bakhuis2025exploring,neaimeh2025learning}. These latency and reliability ranges are indicative design targets derived from grid operation constraints, the IMEP/COS pilot architecture, and existing 5G smart grid literature, rather than empirically validated bounds from a single deployment.

These indicative targets operationalize the domain-level implications from Section~\ref{sec:framework}: business and governance domains drive the need for reliable, auditable market and settlement communication; social concerns drive requirements for user-centric latency and privacy; and infrastructure constraints determine response times for grid-support services.

\subsection{V2G-specific network slicing}
\label{subsec:slicing}

While generic V2X systems are often designed around safety-critical slices for cooperative awareness messages, V2G verticals require a different combination of slices aligned with energy-market and grid-code requirements~\cite{islam2022state,garofalaki2022electric}. Building on the service decomposition in Table~\ref{tab:qos}, we identify three conceptual slice profiles for V2G:

\begin{itemize}
  \item \textbf{V2G control slice}: Carries latency-critical control traffic for fast grid support and distribution-level congestion management, including EV and charger telemetry, set-point updates, and protection-related messages. This slice follows URLLC-like characteristics, with sub-20~ms latency targets in architectures such as the Dresden private-5G deployment~\cite{Wang2026:Bridging}, and must remain operational under grid contingencies.
  \item \textbf{V2G market and settlement slice}: Transports metering data, settlement records, and regulatory reporting information between EVs, aggregators, DSOs/TSOs, and market platforms. Latency constraints are looser (tens of seconds), but integrity, non-repudiation, and long-term retention are critical, reflecting governance and business-domain constraints on auditability and taxation~\cite{wang2026nontechnical,sabadini2025does}.
  \item \textbf{User-facing slice}: Supports mobile applications, dashboards, and notification channels through which users configure participation profiles, provide consent, and receive explanations about control actions. This slice has moderate performance requirements but must enforce strong privacy and access-control policies consistent with social-domain concerns about data exposure and control loss~\cite{bakhuis2025exploring,razzaque2025cybersecurity}.
\end{itemize}

In contrast to generic V2X slices, which primarily optimize for safety and traffic efficiency, V2G slices embed application-layer semantics related to energy contracts, tariffs, and regulatory obligations. The Business/Economic and Governance/Policy domains justify dedicated treatment of settlement and reporting flows, while Social and Infrastructure \& Ecosystem domains motivate the separation of user-facing traffic and latency-critical grid-support traffic, respectively~\cite{wang2026nontechnical}.

\subsection{Edge--cloud partitioning for flexibility control}
\label{subsec:edge-cloud}

The IMEP/COS architecture deployed in Dresden provides a concrete example of how flexibility control can be partitioned across edge and cloud resources~\cite{Wang2026:Bridging}. Three decision loops can be distinguished:

\begin{enumerate}
  \item \textbf{Local edge loop (charger or depot level)}: COS instances at the edge manage charging and discharging decisions based on local state-of-charge (SoC), user constraints (arrival/departure times, minimum SoC), and local grid conditions. These controllers operate at sub-second to second timescales, rely on low-latency communication with wall boxes and building systems, and must provide safe fallback behaviour when backhaul connectivity to IMEP is degraded.
  \item \textbf{Aggregated edge-cloud loop (IMEP)}: The IMEP aggregates charging requests, weather forecasts, energy prices, and building load data, and runs optimization algorithms to generate charging plans on the order of seconds to minutes~\cite{Wang2026:Bridging}. This loop depends on reliable, bandwidth-sufficient connections from distributed edge sites and must respect privacy constraints by working with aggregated or pseudonymized data where required by social-domain considerations~\cite{wang2026nontechnical}.
  \item \textbf{Back-office and market loop}: Central systems interface with electricity markets and regulatory platforms, performing day-ahead and intraday scheduling, settlement, and reporting on timescales of 15~minutes and above~\cite{wang2026nontechnical}. Communication in this loop is less latency-sensitive but must guarantee data integrity, non-repudiation, and consistency with market and grid-code rules.
\end{enumerate}

Each loop is shaped by non-technical domains: business models and regulatory frameworks determine which decisions must be auditable at the back-office; social and privacy requirements constrain which user-level data can leave the edge; and infrastructure maturity influences where real-time observability of grid conditions is available~\cite{wang2026nontechnical}. Algorithmically, these constraints favour rule-based or model-predictive control at the local edge, stochastic or reinforcement-learning-based optimization in the IMEP, and planning and forecasting models in the back-office, all operating under the communication constraints in Table~\ref{tab:qos}.

\subsection{Standards gaps and research directions}
\label{subsec:standards}

Existing 5G systems provide a rich set of QoS classes and network-slicing mechanisms, but they remain largely agnostic to the application-layer semantics of V2G services. In particular, current deployments typically do not expose charging-session states or energy-market events to the 5G core's policy and charging control functions, nor do they natively integrate EV-specific protocols such as ISO~15118 and OCPP into QoS and charging decisions~\cite{garofalaki2022electric,shahed2024battery,Wang2026:Bridging}. As a result, V2G applications must implement their own mapping from non-technical constraints to communication behaviour, rather than relying on vertical-aware network functions.

The socio-technical analysis presented here suggests several research directions. First, network exposure functions and policy control mechanisms could be extended so that aggregators and DSOs can request V2G-specific QoS adaptations (e.g., temporary elevation of control-slice priorities during critical grid events) while respecting regulatory and fairness constraints. Second, deeper integration of ISO~15118/OCPP state machines with 5G core functions would allow charging-session events to trigger slice selection and QoS reconfiguration automatically, aligning communication behaviour with business and governance requirements. Third, future 6G architectures should consider how non-technical domains---such as taxation, user consent, and ecosystem coordination---can be encoded into policy models, enabling communication systems to act as first-class enablers of V2G transition pathways rather than neutral bit pipes~\cite{wang2026nontechnical,Wang2026:Bridging}.

Taken together, these communication requirements and architectural patterns demonstrate that V2G cannot be treated as a generic V2X application. Instead, it demands vertical-specific treatment in 5G system design, with QoS classes, slicing strategies, and edge--cloud partitioning explicitly aligned with the business, governance, social, and infrastructure constraints identified in the underlying socio-technical literature.

\section{Conclusions}
\label{sec:conclusion}

This work derives communication requirements and architectural patterns from an implementation-focused literature corpus and a single urban pilot architecture, rather than from large-scale multi-region deployments. The corpus covers studies up to 2025. Future work will extend the analysis as additional V2G pilots and 5G/6G implementations become available and will empirically validate the proposed QoS classes and slicing patterns in the Dresden Ostra pilot and other MOBILITIES for EU demonstrators.

This article has examined how non-technical factors identified in a systematic review of 974 V2G/V2X studies (2009--2025) translate into communication-system requirements for V2G as a 5G smart grid vertical. Building on the four-domain framework of Business/Economic, Governance/Policy, Social, and Infrastructure \& Ecosystem, the analysis shows how temporal and regional differences in business models, regulation, user practices, and infrastructure maturity impose distinct demands on metering and settlement support, protocol interoperability, privacy-by-design data governance, and latency- and reliability-aware connectivity.

For the communications and networking community, the results clarify that protocol choices, QoS classes, and edge--cloud partitioning for flexibility control cannot be designed in isolation from socio-technical constraints. Instead, V2G communication architectures must support fine-grained metering and settlement, compliance with evolving standards such as ISO~15118 and OCPP, user-centric consent and control mechanisms, and differentiated connectivity for ultra-fast ancillary services versus slower demand-response and planning functions. Treating V2G as a vertical thus shifts the design focus from generic V2X optimization to service-specific requirements aligned with energy markets and grid codes.

Future work includes embedding these non-technical constraints into concrete control and planning frameworks—for example, via digital-twin platforms and machine-learning-based optimization that respect privacy and regulatory limits—as well as validating V2G-specific QoS, security, and data-governance mechanisms in real pilots and testbeds. By aligning 5G and emerging 6G system design with business, governance, social, and infrastructure realities, the communications community can help move V2G from isolated pilots to scalable, sustainable deployment.

\section*{Acknowledgement}
This work is funded by the European Union under Grant Agreement No 101139666, MOBILITIES FOR EU. Views and opinions expressed are those of the authors only and do not necessarily reflect those of the European Union or the European Climate, Infrastructure and Environment Executive Agency (CINEA). Neither the European Union nor the granting authority can be held responsible for them. It is supported by the German Research Foundation (DFG) as part of Germany's Excellence Strategy—EXC 2050/2—Cluster of Excellence “Centre for Tactile Internet with Human-in-the-Loop” (CeTI) of Technische Universität Dresden under project ID 390696704 and the Federal Ministry of Research, Technology, and Space (BMFTR) for its support as part of the research program Communication Systems “Souverän. Digital. Vernetzt.”. Joint project 6G-life, project identification number: 16KIS2413K.

\bibliographystyle{IEEEtran}
\bibliography{references}
\end{document}